\documentclass[prl,showpacs,floatfix,twocolumn]{revtex4}
\usepackage{graphicx}
\usepackage{dcolumn}

\begin{document}

\newcommand*{\cm}{cm$^{-1}$\,}

%
\title{Superconducting energy gap versus pseudogap in hole-doped cuprates as revealed by infrared spectroscopy}
%
%
\author{T. Dong}
\author{F. Zhou}
\author{N. L. Wang}

\affiliation{Beijing National Laboratory for Condensed Matter
Physics, Institute of Physics, Chinese Academy of Sciences,
Beijing 100190, P. R. China}
%
%

\begin{abstract}

We present in-plane infrared reflectance measurement on two superconducting cuprates with relatively low T$_c$: a nearly
optimally-doped Bi$_2$Sr$_{1.6}$La$_{0.4}$CuO$_{6+\delta}$ with T$_c$=33 K and an underdoped
La$_{1.89}$Sr$_{0.11}$CuO$_4$ with T$_c$=30 K. The measurement clearly reveals that the superconducting energy gap is distinct from the pseudogap.
They have different energy scales and appear at different temperatures. The results suggest
that the pseudogap is not a precursor to the superconducting
state. The data also challenge the longstanding viewpoint that the superconductivity within the ab-plane is in the clean limit and the
superconducting pairing energy gap could not be detected by in-plane infrared spectroscopy.
\end{abstract}


\maketitle

%

The energy gap created by the pairing of electrons is the
most important parameter of a superconductor. Probing the pairing
energy gap is crucial for elucidating the mechanism of
superconductivity. For conventional superconductors, infrared
spectroscopy is a standard technique to probe the superconducting
energy gap, as the electromagnetic radiation below the gap energy
2$\Delta$ could not be absorbed.\cite{Tinkham} For high
temperature superconductors (HTSC), however, the situation is
rather unclear. A predominant view is that the superconducting energy gap
could not be detected from the ab-plane infrared spectra because
the HTSCs are in the clean limit.\cite{Kamaras} Although some
features were actually seen in the low-frequency reflectance and
conductivity spectra, they were widely ascribed to either the
onset of mid-infrared component or the coupling effect of
electrons with some bosonic excitations.

Another reason that complicates the identification of the
superconducting energy gap is the presence of pseudogap. The pseudogap was observed for almost all
underdoped high-T$_c$ cuprates,\cite{Timusk} and many optimally
doped systems including the most commonly studied
Bi$_2$Sr$_2$CaCu$_2$O$_{8+\delta}$ (Bi2212)\cite{Lee}, and
Bi$_2$Sr$_2$CuO$_{6+\delta}$ (Bi2201)\cite{Kondo}. Early
angle-resolved photoemission (ARPES)\cite{Ding,Loeser,Kampuzano}
and scanning tunneling microscopy (STM)
experiments\cite{Renner,Kugler} on underdoped samples indicated
that the superconducting gap smoothly evolves into the pseudogap
state with increasing temperature. The lack of any obvious change
at T$_c$ for the gap amplitude has been taken as important
evidence for the one gap picture that the pseudogap is a precursor
to the superconducting state but lacks its pairing phase
coherence. However, in recent years, several ARPES measurements on underdoped
Bi2212,\cite{Tanaka,Lee} optimally doped Bi2201\cite{Kondo} and
La$_{2-x}$Sr$_x$CuO$_4$ (LSCO)\cite{Terashima}, as well as
Raman\cite{Tacon} and STM\cite{Boyer} studies, revealed a second
energy gap forming abruptly at T$_c$ on the Fermi arc near nodal
region. This gap has a canonical BCS-like temperature dependence
and is accompanied by the appearance of the Bogoliubov
quasi-particles.\cite{Lee} So it represents the order parameter of
superconducting state, whereas the pseudogap near the antinodal
region is an energy scale associated with a different mechanism. Recently, a number of experimental investigations
indicated that the pseudogap is associated with the charge-density-wave order and it competes with the superconductivity
\cite{Ghiringhelli,Chang,Blackburn,Blanco-Canosa,Wu,LeBoeuf}.
It is noted that the superconducting gaps are close or comparable
to the pseudogaps for systems with relatively higher T$_c$s (for
example, in not heavily underdoped Bi2212 or
YBa$_2$Cu$_3$O$_{6+\delta}$ (YBCO)). On the other hand, for
optimally doped Bi2201,\cite{Kondo} LSCO,\cite{Terashima} or
heavily underdoped Bi2212,\cite{Tanaka} the gaps formed at the
Fermi arc, including their simple extrapolation to the antinodal
position, are significantly smaller than the antinodal pseudogaps.

To avoid possible complication arising from similar gap amplitudes
between the superconducting gap and the pseudogap, here we study
two different systems with relatively lower T$_c$: a nearly
optimally-doped Bi$_2$Sr$_{1.6}$La$_{0.4}$CuO$_{6+\delta}$
(La-doped Bi2201 with T$_c$=33 K) and an underdoped
La$_{1.89}$Sr$_{0.11}$CuO$_4$ (LSCO, T$_c$=30 K). We
observed an abrupt spectral change at low frequency directly
associated with the superconducting transition in both
cuprate systems. We elucidate that those changes are caused by the
formation of d-wave superconducting gap below T$_c$. At higher
frequencies, another shoulder feature is present in reflectance
and shows little change across T$_c$. It is caused by the partial
energy gap in the Fermi surface. Our study reveals that the superconducting gap and pseudogap are two distinct energy gaps. They
have different energy scales and appear at different temperatures. The work enables us to reconcile the optical spectroscopy
result with other experimental measurements.

High quality single crystals in both systems were grown by floating zone method \cite{Zhou}. The near-normal
incident reflectance were measured using both Bruker 66v/s and
113v spectrometers with \textit{in-situ} overcoating technique.
The optical conductivity was obtained by performing Kramers-Kronig
transformation.

\begin{figure}[t]
\centerline{\includegraphics[width=2.5in]{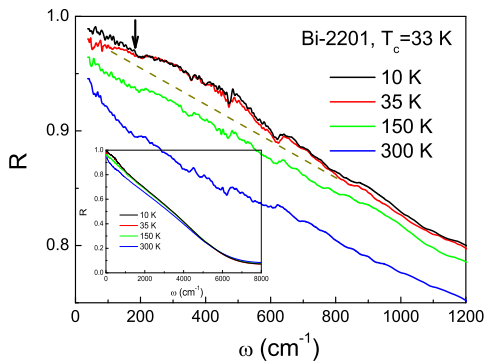}}%
\vspace*{-0.20cm}%
\caption{(Color online) The temperature-dependent reflectance
R($\omega$) for Bi2201 below 1200 \cm. An upward deviation from
linear-$\omega$ dependence below roughly 800 \cm is seen in
R($\omega$) at 10 K and 35 K. The dashed straight line is a guide
for the eyes. This gives a weak shoulder at around 400-800 \cm in
R($\omega$). Below T$_c$, a further upturn is observed at lower
frequency as indicated by an arrow. The inset shows the data taken over broad frequencies.}%
\label{fig1}
\end{figure}

Let us first look at the data collected on La-doped Bi2201
crystal. Figure 1 shows the temperature-dependent reflectance
R($\omega$) below 1200 \cm. The inset shows the data taken over
broad frequencies. We notice that R($\omega$) roughly displays the
well-known linear-$\omega$ dependence over broad frequencies at
high temperatures, \textit{i.e.} 300 K or 150 K. This gives rise
to the approximately linear-frequency dependent optical scattering
rate, as shown in Fig. 2 (a), in terms of the extended Drude model
1/$\tau$(${\omega}$)=$(\omega_p^2 / 4\pi)$Re[1/$(\sigma(\omega)]$
where $\omega_p$ is the overall plasma frequency and can be
obtained by summarizing optical conductivity up to the reflectance
edge frequency. Its lineshape appears just like an up-side-down of
the R($\omega$). However, at low temperatures, \textit{e.g.} at 10
K and 35 K, the R($\omega$) curves apparently deviate upward from
the linear-$\omega$ dependence below roughly 800 \cm. A curvature
is seen very clearly for the R($\omega$) curves at the main panel.
The dashed straight line is a guide for the eyes. This curvature
is essentially the same as the prominent shoulder features
observed in other systems with relatively higher T$_c$s, such as
YBCO\cite{Hwang1}, Bi2212\cite{Hwang2} or Tl-based
systems\cite{Ma}, although much weaker here. In the
1/$\tau$(${\omega}$) spectrum, the low temperature scattering rate
shows a downward suppression at low frequencies. Those spectral
features were taken as the optical signature of pseudogap state,
but later were frequently assigned to the coupling effect of
electrons with certain bosonic mode\cite{Hwang1,Hwang2}. According
to previous optical studies on the electronic systems with partial
energy gaps formed on the Fermi surface, for example, in 2D
transitional metal dichalcogenide CDW system\cite{Hu} or
electron-doped cuprate system Nd$_{2-x}$Ce$_x$CuO$_4$\cite{Wang},
such spectral structures could be unambiguously caused by the
partial energy gaps.

\begin{figure}
\centerline{\includegraphics[width=2.5in]{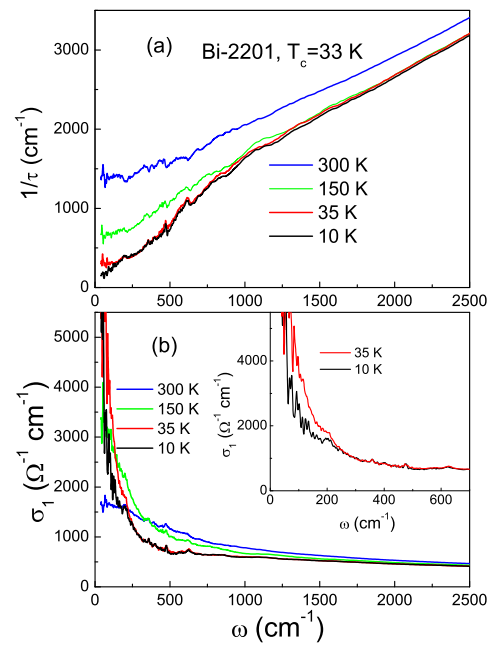}}%
\vspace*{-0.20cm}%
\caption{(Color online) (a) The scattering rate, and (b) the
optical conductivity spectra for Bi2201 at different temperatures.
This inset shows the conductivity spectra in the superconducting
state and the normal state just above T$_c$.}%
\label{fig1}
\end{figure}

The most important observation in this work is that a further
spectral change occurs below superconducting transition. The
reflectance at 10 K below 200 \cm shows a clear further upturn
from the R($\omega$) curve at 35 K (above T$_c$). This spectral
change was repeatedly observed in different pieces of crystals
from the same crystal rod. Similar spectral change is also seen in
underdoped LSCO below superconducting transitions as we
shall present below. Thus it represents a new energy scale
associated with the superconducting transition. The spectral change is not significant and the low-$\omega$ reflectance at
the temperature far below T$_c$ does not approach unit abruptly. This could be attributed to the
d-wave energy gap. The low energy quasiparticle excitations 
are still present due to the presence of nodes. It deserves to
remark that, in earlier optical studies on underdoped
high-temperature superconductors with relatively higher T$_c$,
including commonly studied Bi2212\cite{Hwang2},
YBa$_2$Cu$_3$O$_{7-\delta}$\cite{Hwang1},
YBa$_2$Cu$_4$O$_8$\cite{Basov}, this second abrupt spectral change
below T$_c$ was not observed. In those systems, no qualitative
difference in the spectra between the pseudogap state and the
superconducting state was seen in infrared experiment, similar to
the observations in earlier ARPES\cite{Ding,Loeser,Kampuzano} and
tunneling measurements\cite{Renner,Kugler}. This leaded to the
conclusion that the pseudogap state was already a lot like the
superconducting state. We note that this statement is only true
for the spectra taken above 200 \cm at 10 K and 35 K here,
suggesting that the spectral structure related to the pseudogap
energy does not change across the T$_c$; on the contrary, the
spectral change in R($\omega$) below 200 \cm is directly caused by
the d-wave superconducting pairing. It leads to a reduction of the
spectral weight in optical conductivity at very low energies, as
shown in Fig. 2 (b). The missing spectral weight is transferred to
the strength of delta function at zero frequency, representing the
superconducting condensate.

It is worthwhile to compare the optical data with the result
obtained from the ARPES measurement on similar La-doped Bi2201
crystal with approximately the same T$_c$\cite{Kondo}.
ARPES study clearly revealed the existence of a gapless Fermi arc
near the nodal region and an energy gap about 40 meV near the
antinodal region ($\pi$, 0) above T$_c$ (but below the pseudogap
closing temperature T$_{PG}$). The antinodal energy gap does not
show much difference as the sample becomes superconducting,
however, a second energy gap opens up on the Fermi arc only below
T$_c$. Its energy scale is about 10-15 meV, being distinct from
the magnitude of the pseudogap.\cite{Kondo} We find that our
optical data are in very good agreement with ARPES experiment,
considering that the optical gap should double the ARPES
measurement being relative to the Fermi energy. The weak shoulder
in R($\omega$) between 600-800 \cm is associated with the partial
energy gap, \textit{i.e.} the pseudogap, near the antinodal
region, while the new energy scale below 200 \cm is associated
with the d-wave pairing gap opened up on the nodal Fermi arc.
Above 150 K, the spectral feature linked with pseudogap could not
be well resolved in our infrared data, this is also consistent
with the ARPES measurement that the pseudogap is already closed at
150 K.\cite{Kondo} Our experiment strongly suggests that the
spectral change caused by the pairing gap below T$_c$ could be
detected from the infrared spectroscopy.

\begin{figure}
\centerline{\includegraphics[width=2.5in]{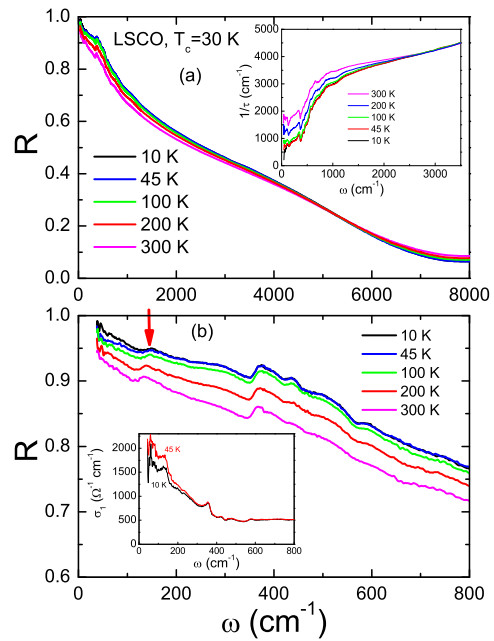}}%
\vspace*{-0.20cm}%
\caption{(Color online) The temperature-dependent reflectance
R($\omega$) for La$_{1.89}$Sr$_{0.11}$CuO$_4$ (a) over broad
frequencies, and (b) below 800 \cm. The arrow indicates the
frequency below which a further upturn appears below T$_c$. The
insets show the scattering rate
and the conductivity spectra above and below T$_c$, respectively.}%
\label{fig1}
\end{figure}

Figure 3 shows the measured in-plane reflectance data for the
underdoped La$_{1.89}$Sr$_{0.11}$CuO$_4$ crystal with T$_c$=30 K:
(a) the data over broad frequencies up to 8000 \cm, (b) the data
at low frequencies, below 800 \cm. As the sample is rather
underdoped, the reflectance does not show a linear-frequency
dependence. A pronounced shoulder is seen near 500-700 \cm for
spectra at all measuremed temperatures. The reversed S-like shape
is a strong indication for the presence of a partial energy gap in
the Fermi surface.\cite{Wang} In the scattering rate spectra shown
in the inset of Fig 3 (a), the strong suppression below 700 \cm is
seen for all curves. Like the case of La-doped Bi2201, those
features could be ascribed to the partial energy gaps at antinodal
region which should be persistent even above room temperature in LSCO. In
the expanded plot of R($\omega$) at low frequencies (Fig. 3 (b)),
a further upturn is observed in curve at 10 K only below 150 \cm
from the normal state R($\omega$) at 45 K. Similar to La-doped
Bi2201, this spectral change is linked with the superconducting
gap below T$_c$. It causes a small missing area at low frequencies
in optical conductivity, as shown in the inset of Fig. 3 (b).

Summarizing our infrared measurement on two different
superconducting systems with relatively lower T$_c$, we find two
major structures in optical spectra. One is a shoulder feature at
relatively higher energy scale in R($\omega$), roughly 600-800
\cm. The feature is rather weak in Bi2201 sample, and not visible
above 150 K. In underdoped LSCO, the feature is strong,
and persistent above room temperature. The other one, which is
more important and not resolved in earlier optical studies on
systems with relatively higher T$_c$, is the identification of a
second energy scale about 150-200 \cm directly associated with the
superconducting transition.

It is highly interesting to discuss the origin of the second
energy scale which is directly associated with the superconducting
transition. Although it is very natural to assign it to the
formation of the superconducting energy gap, one may argue that
this kind of spectral change may originate from the coupling
effect of electrons with certain bosonic mode.\cite{Hwang1,Hwang2}
Let us discuss this possibility first. In high-T$_c$ cuprates, two
candidates for sharp bosonic mode could exist: phonon and magnetic
excitation (with a resonance at ($\pi, \pi$)). Since the phonon
mode could not disappear suddenly above T$_c\sim$30 K,
furthermore, the frequency is already much lower than any known
phonon mode involved in the in-plane Cu-O vibrations, phonon mode
could be ruled out. As for the magnetic resonance mode, neutron
studies on bilayer cuprates YBCO and BSCCO revealed that the mode
occurs only below T$_c$ at optimal doping. For underdoped samples,
broad mode feature could be observed above T$_c$, but it locates
at the same energy scale as below T$_c$.\cite{Dai} Here the LSCO
crystal is substantially underdoped, however, the
feature is only observed below T$_c$. Additionally, the magnetic
resonance at ($\pi, \pi$) was not observed in the single-layered
compound. Therefore it is very unlikely that the feature is linked
with magnetic excitations. Then, we are left with the sole
possibility, that is, the gap formation caused by the
supercondcuting pairing.

Our experiment severely challenges the point of view that the
superconducting gap in high-T$_c$ cuprates could not be observed
in infrared spectroscopy. Such statement was made based on the
assumption that the superconductivity in ab-plane was in the clean
limit.\cite{Kamaras} In this case, the quasiparticle
mean-free-path is much longer than the coherence length
\textit{l}$\gg\xi$, or equivalently the normal-state scattering
rate is much smaller than the superconducting gap amplitude
$1/\tau\ll\Delta$. However, this clean limit scenario is under
intense debate.\cite{Homes1} In earlier studies on this issue, the
anisotropic nature of the gap, the scattering rate and the Fermi
velocity, were not sufficiently considered. Quite often, the
average value of the mean-free-path or the value near the nodal
region was used to compare with the coherence length near
antinodal region. We argue that this clean limit criteria could
not be fulfilled in the d-wave superconductivity in cuprates.
In the nodal region, although the scattering rate is small, the
gap amplitude is also very small (it is virtually zero at the
nodal point, thus leading to divergence of the coherence length).
In the antinodal region, the gap is large, but the quasiparticles
experience very strong scattering, or even could not be
well-defined in the underdoped case. Thus, generally the
clean-limit criteria $1/\tau\ll\Delta$ could not be fulfilled. On
this basis, the pairing gap is expected to be observed by infrared
spectroscopy. That is what we find in this work. We note that in
electron-type cuprates as well as in certain composition of La$_{2-x}$Ba$_x$CuO$_4$, the superconducting gap was also
observed.\cite{Zimmers,Homes2,Homes3}

\begin{figure}[t]
\centerline{\includegraphics[width=3.2in]{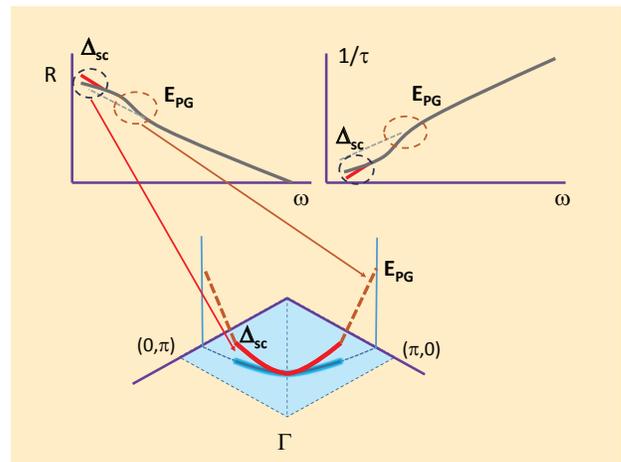}}%
\vspace*{-0.20cm}%
\caption{(Color online) A schematic picture showing the relation
between the gap features in infrared reflectance spectrum and the two
distinct gaps seen in ARPES.}%
\label{fig1}
\end{figure}

The comparison between our data and ARPES as we presented above
strongly suggests that low-frequency spectral change in
R($\omega$) below T$_c$ probes the superconducting gap formed on
the Fermi arc near the nodal region, while the shoulder feature in
R($\omega$) at higher frequencies is associated with the antinodal
gap near ($\pi$, 0). A schematic picture for the relation between
the structures seen in infrared spectrum and ARPES is shown in
Fig. 5. We note that our experiment is not consistent with the
one-gap scenario that the pseudogap is a precursor to the
superconducting state. On the contrary, our work provides an
optical evidence for two energy gaps for the superconducting
state. It supports the picture that the gap near the antinodal
region is associated with the non-superconducting order parameter, e.g. the CDW order as evidenced by
a number of recent experimental probes \cite{Ghiringhelli,Chang,Blackburn,Blanco-Canosa,Wu,LeBoeuf},
while the gap which opens on the Fermi arc is associated with the
superconductivity. The present work enables us to reconcile the optical spectroscopy
probe with other experimental measurements.

This work was supported by the National Science Foundation of
China, and the 973 project of the Ministry of Science and Technology of China (2011CB921701,2012CB821403).

%
%

\end{document}